# Analysis of InAs Vertical and Lateral Band-to-Band Tunneling Transistors: Leveraging Vertical Tunneling for Improved Performance

Kartik Ganapathi, Youngki Yoon and Sayeef Salahuddin[a]

Department of Electrical Engineering and Computer Sciences, University of California Berkeley, CA – 94720

## ABSTRACT

Using self-consistent quantum transport simulation on realistic devices, we show that InAs band-to-band Tunneling Field Effect Transistors (TFET) with a heavily doped pocket in the gate-source overlap region can offer larger ON current and steeper subthreshold swing as compared to conventional tunneling transistors. This is due to an additional tunneling contribution to current stemming from band overlap along the body thickness. However, a critical thickness is necessary to obtain this advantage derived from 'vertical' tunneling. In addition, in ultra small InAs TFET devices, the subthreshold swing could be severely affected by direct source-to-drain tunneling through the body.

a) sayeef@eecs.berkeley.edu

The recent interest in Band-to-Band Tunneling Field Effect Transistors (TFETs) is due to the promise of overcoming the fundamental limit of subthreshold swing (60 mV/decade at room temperature) in case of classical MOS devices, thereby providing a path to significantly reduce supply voltage and power dissipation. The advantage of TFETs over MOSFETs in this regard has been discussed and demonstrated in various material systems.[1-3] However, the ON state current in TFETs is significantly lower than in MOSFETs owing to tunneling resistance which arises due to carrier tunneling between eigenstates of different symmetry (i.e. conduction and valence band). Hence it is critical to the design of TFETs to understand the factors governing ON current and subthreshold swing. In this letter, we investigate the performance of InAs TFETs where tunneling occurs in the direction normal to the semiconductor-dielectric interface (vertical TFET from hereon) to see if there are any inherent advantages owing to either the device geometry or the nature of tunneling in these devices over conventional TFETs where the tunneling occurs from source to drain (lateral TFET from hereon).[4-5] To this end, we perform self-consistent ballistic quantum mechanical simulations with multi-band Hamiltonian within the Non-Equilibrium Green's Function (NEGF) formalism in realistic-size devices. Our simulations show that vertical TFETs, due to an additional vertical tunneling component provide more ON current than their lateral counterparts. We also explore the design possibilities of vertical TFETs whereby we show that they can be optimized to yield steeper turn-on characteristics and smaller OFF currents than in lateral TFETs. Our results also provide insight towards scaling (in terms of both body thickness and channel length) behavior of ultra-thin body vertical TFETs based on low bandgap and low effective mass materials.

Figure 1 shows the schematic of the cross section of simulated devices where the lateral TFET [Fig. 1(a)] is an ultra thin body double-gate InAs p-i-n TFET while the vertical InAs TFET [Fig. 1(b)] is a device with identical footprint except for a heavily doped n+ pocket in the gate-source overlap region. Our vertical TFET has an additional back-gate underneath the channel in comparison to the device structure proposed by Pratik et al. (see Fig. 2 of Ref. 5) in order to have a better electrostatic control of while keeping the vertical tunneling intact. The doping densities, device dimensions and other parameters used in simulation are mentioned in the caption of Fig. 1. A 4 × 4 *k.p* Kane's second order Hamiltonian is used to describe the bandstructure of InAs.[6] The spurious states in the dispersion curves arising due to confinement are removed following standard techniques.[7] Spin-orbit coupling has been ignored in our simulations to reduce computation time.[8]

The Green's function *G*, at a given total energy *E*, calculated using the self-consistent Born approximation is given by $G(E) = [EI – H - \Sigma_1 - \Sigma_2]^{-1}$ where *H* is the Hamiltonian of the system, *I* an identity matrix and $\Sigma_{1,2}$ the contact self-energies.[9] The contact self-energies are calculated by assuming semi-infinite leads using a technique due to Sancho et al. [10], while *G* is calculated using the recursive Green's function algorithm.[11] We assume that the dimension of the devices is large along the width and hence use a periodic boundary condition in that direction. We note that while using a multi-band Hamiltonian, it is not possible to analytically sum over the momentum states along the width in the calculation of electron and hole densities as can be done in case of single band Hamiltonians.[12] Therefore the transverse modes are considered numerically. The electron and hole densities along with the electrostatic potential $U$ are computed self-consistently by iteratively solving Pöisson and NEGF equations. The current is then calculated using the converged potential profile.[9] The transport simulation is parallelized. A typical iteration for 10 nm thick and 60 nm long devices takes around 25 seconds using 16000 cores in parallel. Traditionally, a full NEGF simulation of realistic devices has been prohibitive due to computational burden and most simulations are based on unreasonably small approximation of actual devices. In our case, massive parallelization enables us to solve for realistic structures.

Figure 2(a) shows the plot of drain current $I_D$ versus gate voltage $V_G$ at a drain voltage $V_D$ of 0.4 V for lateral and vertical TFETs with channel lengths 20 nm. It can be seen that the vertical TFETs have a smaller OFF state current as compared to their lateral counterparts. The reason for this can be understood by looking at the energy band profiles along the source-drain direction, which, at the semiconductor-dielectric interface are plotted in Fig. 2(b) for gate voltages near the OFF state. The vertical TFET, due to the pocket doping, has an additional tunneling barrier on the source side as compared to lateral TFET, which suppresses the penetration of tunneling states into the channel. This is confirmed by the fact that the difference in OFF state currents is less pronounced in case of 30 nm channel length devices [see OFF currents in Figs. 3(a) and 3(b)].

Another important observation to be made from the $I_D$-$V_G$ characteristics is that the vertical TFET has a steeper subthreshold swing than the lateral device. The explanation for this is twofold. First, the vertical device can be envisioned as a gated p$^+$-n$^+$ diode (source-pocket junction) in series with a MOSFET (pocket-channel junction) with fully depleted source. The potential barrier of the MOSFET is lowered by the gate voltage at a rate similar to that of channel potential in the lateral TFET. However the *tunneling width* for a p$^+$-n$^+$ junction is smaller than that of a lateral TFET. The current is dominated by the extent of the tunneling width once the MOSFET potential barrier is lowered sufficiently and therefore will be larger than that of lateral TFET at similar gate voltages, leading to a steeper subthreshold swing due to smaller OFF current. A second reason for the steeper swing can be attributed to the onset of vertical tunneling in the region underneath the pocket due to band overlap that also contributes to larger current. This is clearly seen from the energy band profiles along the thickness of the device shown in Fig. 2(c). The contribution to current by vertical tunneling continues to increase for larger gate voltages due to increased band overlap along the thickness.

One of the interesting results our studies show is that of the existence of a minimum body thickness below which the vertical tunneling is absent. This can be attributed to two main causes – a) a larger bandgap at smaller body thicknesses b) a fully depleted p$^+$ region underneath the pocket. We simulate a vertical TFET with 6 nm body thickness and a 2.4 nm thick pocket wherein the back-gate is absent in order to ensure that the same does not adversely affect the vertical band bending. The energy band diagrams along the body thickness for such a device, shown in Fig.

2(d) confirms the absence of band overlap due to aforementioned reasons. In comparison, the device with 10 nm body thickness, as shown in Fig. 2(c), clearly shows a band overlap.

Another intriguing feature of the $I_D$-$V_G$ characteristics in Fig. 2(a) is the fact that the subthreshold swing is larger than 60 mV/decade in both lateral and vertical TFETs, contrary to the fact that band-to-band tunneling should provide less than 60 mV/decade. Similar degraded subthreshold swing has been seen previously.[13] We confirmed from our simulations that this is not due to poor electrostatics as our devices have excellent gate control (i.e. $\partial\varphi_s/\partial V_G > 0.96$, $\varphi_s$ being the electrostatic potential at the semiconductor-dielectric interface). To investigate the reasons for this, we study the scaling behavior of vertical TFETs by varying the length of the channel. Figures 3(a) and 3(b) show the $I_D$-$V_G$ characteristics of vertical and lateral TFETs respectively for three different lengths of the channel – 10 nm, 20 nm and 30 nm with all other parameters remaining the same. Evident from the characteristics is the fact that these devices exhibit very poor scalability. Our simulations show that this is mainly due to the fact that the conduction band effective mass in InAs is quite low ($0.03m_0$ for a 10 nm thick body) which leads to huge penetration of wavefunctions into the channel and hence a large leakage current which limits the swing in TFETs. The Local Density of States (LDOS) plot on a logarithmic scale in Fig. 3(c) shows large penetration of tunneling states into the channel in the OFF state. Energy resolved current, $I(E)$ given by $T(E) \times (f_1(E) - f_2(E))$, is plotted in Fig. 3(d). Here $T(E)$ is the transmission probability at energy $E$ and $f_1$ and $f_2$ are the Fermi-Dirac distributions corresponding to source and drain respectively. All the plots correspond to OFF state. From Fig. 3(d), the peak of the current appears in the energy range of 0 to -0.5 eV, where the bandgap in the channel should have stopped the current flow [see Fig. 3(c)]. This then clearly points to direct source-to-drain tunneling that becomes increasingly severe as one goes down to smaller channel lengths.

To summarize, using self-consistent ballistic NEGF simulations on realistic structures, we have shown that the vertical TFETs offer significant advantages over their lateral counterparts in terms of increased ON current and steeper subthreshold swing. This is due to (i) an additional tunneling barrier in the current path at the OFF state that provides lower OFF current, (ii) a thinner tunneling barrier in the ON state that provides larger ON current and (iii) finally, a vertical tunneling component in addition to the lateral one in the ON condition that further increases the drive current. In this study, we have restricted ourselves to using nominal device structures so as to retain our emphasis on the underlying physics. Nonetheless, the aforementioned points indicate ripe opportunities for optimization by band engineering either through strain or through heterostructures and thereby amplify the advantages of vertical TFETs over conventional lateral TFETs. Our simulations also shed light on both the vertical and lateral scaling behavior of ultra-thin body tunnel TFETs based on low bandgap and low effective mass materials.

The authors would like to thank P. Patel and C. Hu for bringing this problem to their attention and also for many useful discussions. This work was in part supported by FCRP center on Functional Engineered and Nano Architectonics

# REFEENCES

**FIGURE CAPTIONS**

Figure 1. Schematic of the simulated devices (a) Lateral InAs TFET (b) Vertical InAs TFET with a heavily doped n+ pocket (halo) in the gate-source overlap region. The doping profiles in all our simulations are abrupt with a source and drain doping of $5 \times 10^{19}$ cm$^{-3}$ and $5 \times 10^{18}$ cm$^{-3}$ respectively. The asymmetry in doping concentrations is motivated by the lower conduction band density of states in InAs and the need to suppress ambipolar conduction. In case of vertical TFET, we use a pocket doping of $5 \times 10^{19}$ cm$^{-3}$. Although the channel region is intrinsic, we use an n-type doping of $1 \times 10^{15}$ cm$^{-3}$ to account for unintentional doping arising due to defects. A 1.2 nm gate dielectric with $\kappa$ = 15.4 is used. The length of source and drain in our simulations is 20 nm each with a 10 nm overlap on the source side in case of vertical TFET. For 10 nm thick body, the pocket is 3.6 nm deep while for a body thickness of 6 nm, we use a pocket depth of 2.4 nm. The crystallographic direction is assumed to be (100) for transport and the body is confined along (001) direction. Also, the difference in workfunction between semiconductor and gate metal - $\Phi_{ms}$ is assumed to be zero.

Figure 2. (a) $I_D$-$V_G$ characteristics at $V_D$ = 0.4 V for lateral and vertical TFETs for channel length $L_{ch}$ = 20 nm. The markers indicate values from simulation and lines, the interpolated curve. (b) The energy band diagrams in a 10 nm thick vertical TFET along the lateral direction near the semiconductor-dielectric interface for different gate voltages – from 0.15 V to 0.45 V in steps of 0.1 V. (c) The energy band diagrams along the body thickness in the middle of the pocket for same gate

voltages as in (b). (d) Similar band diagrams for a 6 nm thick device with no back-gate for two different gate voltages – 0.8 V (blue) and 1.1 V (green). From (a), we can see that the vertical TFETs have smaller OFF state currents and larger ON currents than their lateral counterparts. The vertical TFETs also have steeper subthreshold swing. While (c) shows band overlap for the 10 nm case, (d) does not show any overlap even at high gate voltages (1.1 V). Note that in (d), a single gate geometry has been chosen to maximize the possibility of band overlap as a double gate structure could further degrade the band overlap.

Figure 3. (a) $I_D$-$V_G$ characteristics at $V_D$ = 0.4 V for vertical TFETs for three different channel lengths – 10 nm, 20 nm and 30 nm showing the scaling behavior. (b) Similar curves for lateral TFETs. From (a) and (b), it is evident that TFETs show poor scalability and show less than 60 mV/decade subthreshold slope only for longer channel lengths. (c) Local Density of States (LDOS) on a logarithmic scale for vertical TFET with $L_{ch}$ = 20 nm at $V_G$ = 0.2 V near the semiconductor-dielectric interface showing large penetration of tunneling states in the channel owing to smaller effective mass of InAs. (d) Energy resolved current density for lateral TFET with different channel lengths at $V_G$ = 0.2 V showing significant current flowing due to tunneling through the channel for shorter devices, thereby limiting the maximum subthreshold slope and minimum OFF current.

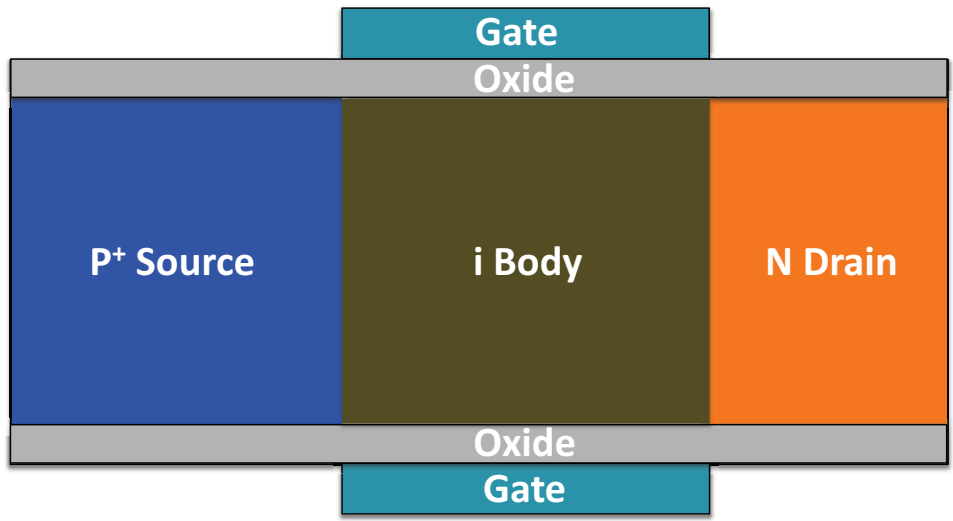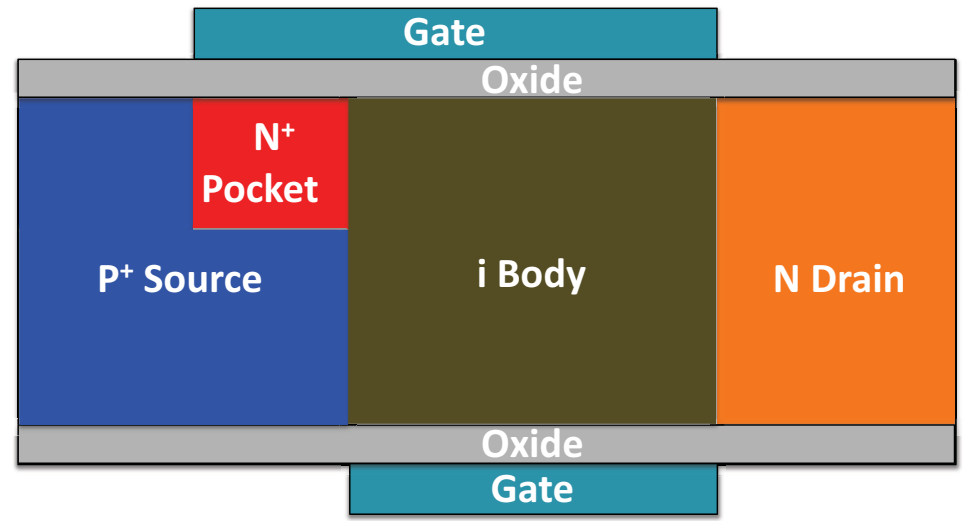

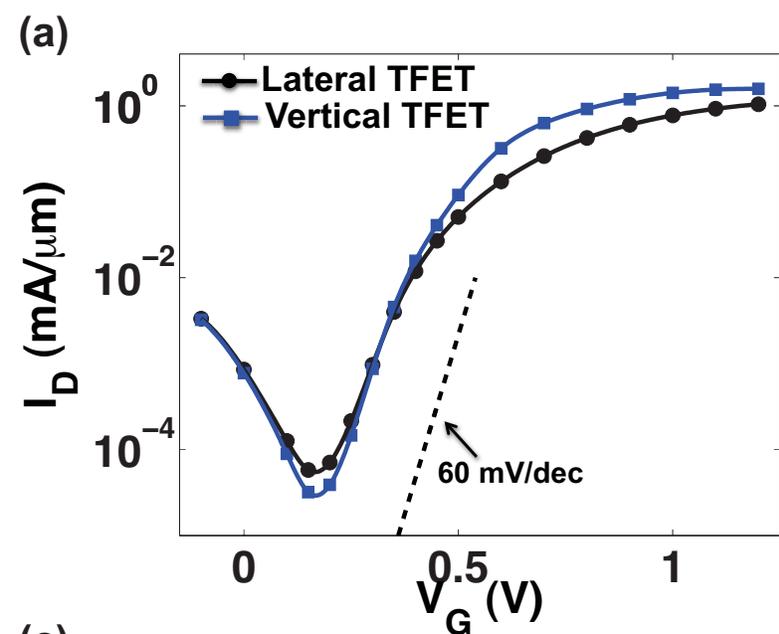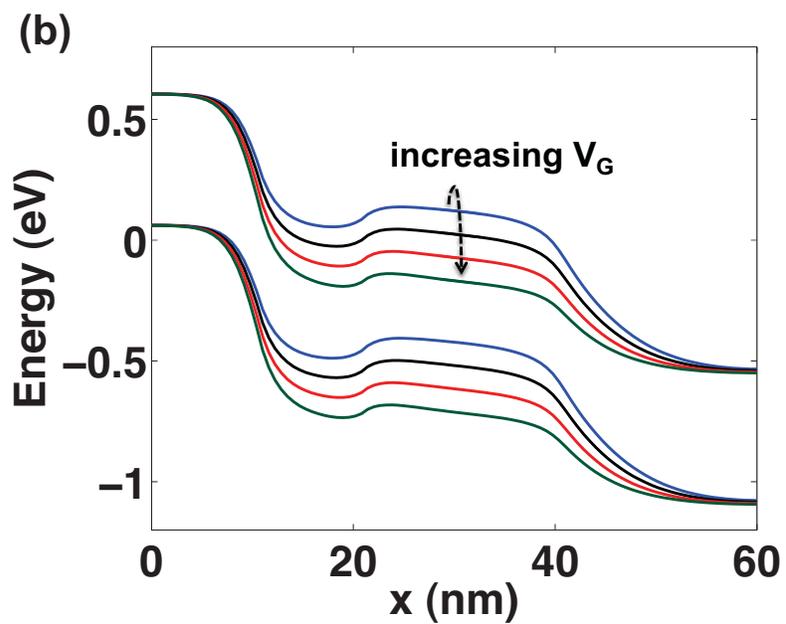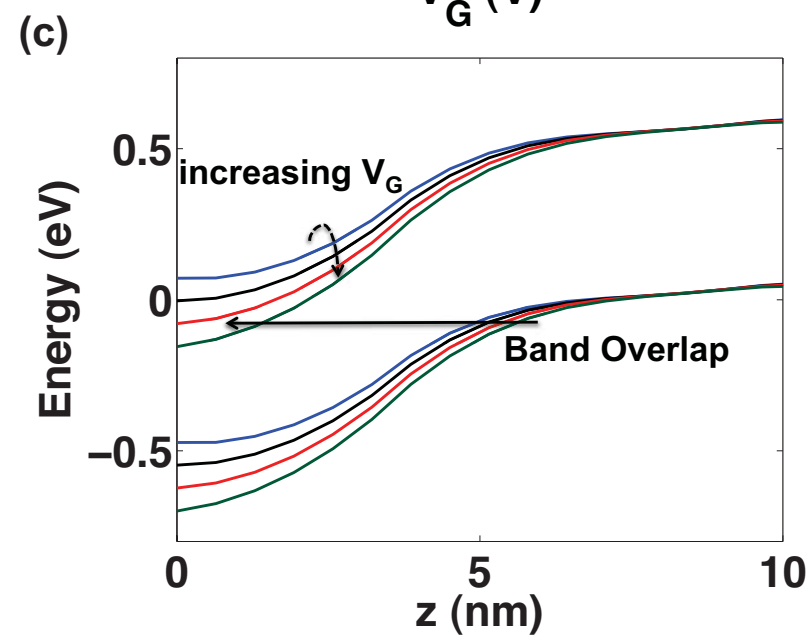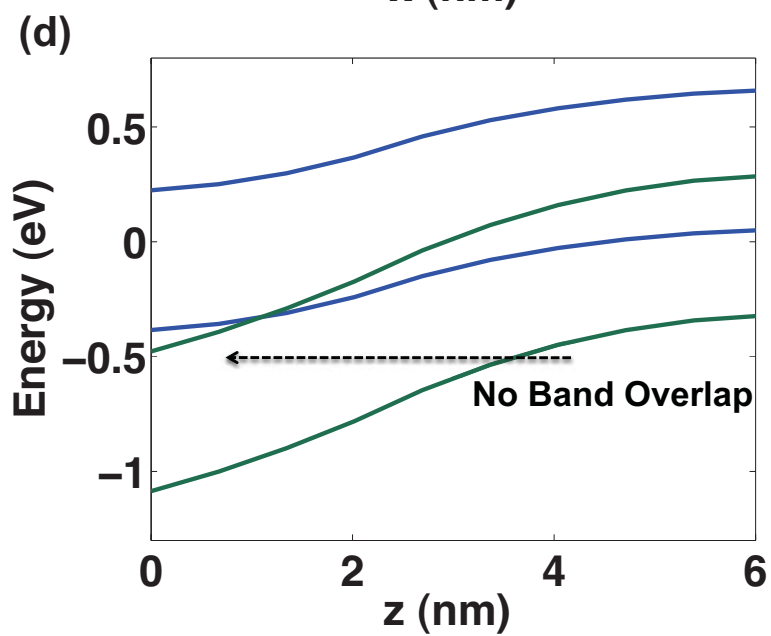

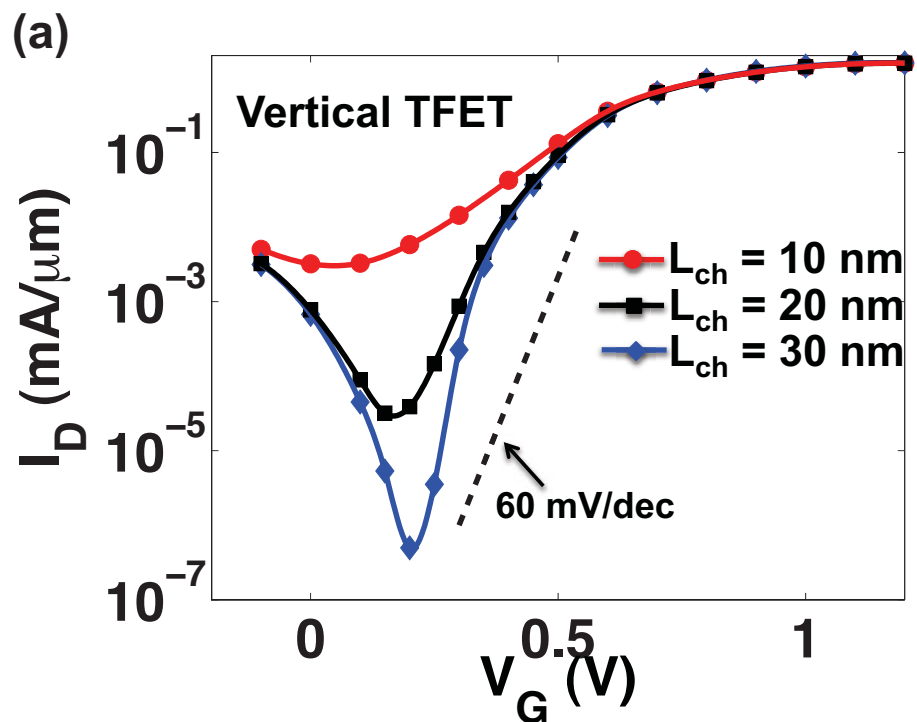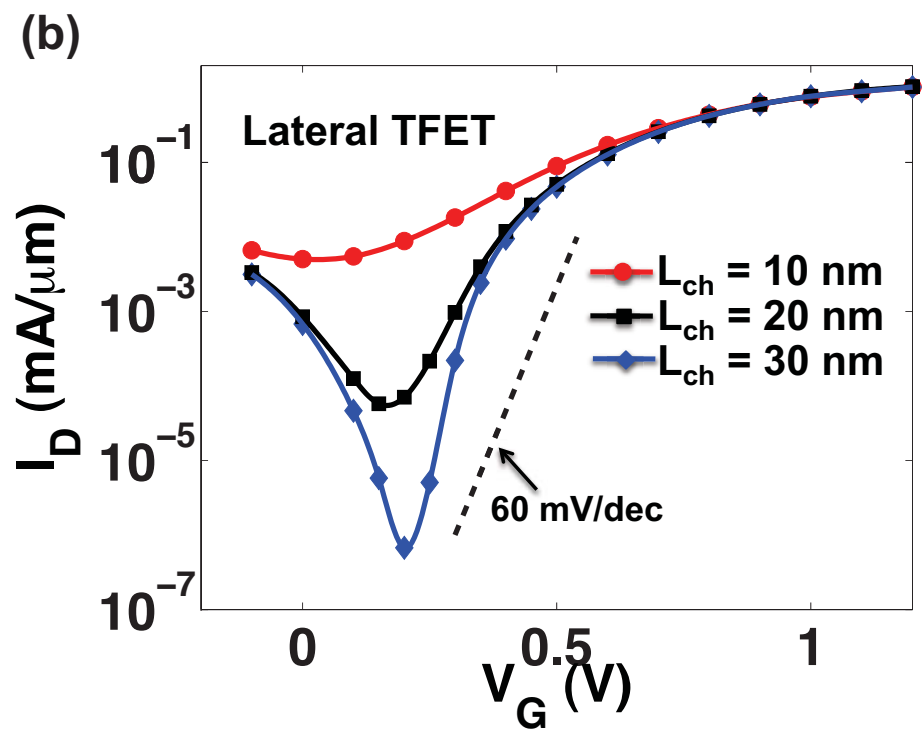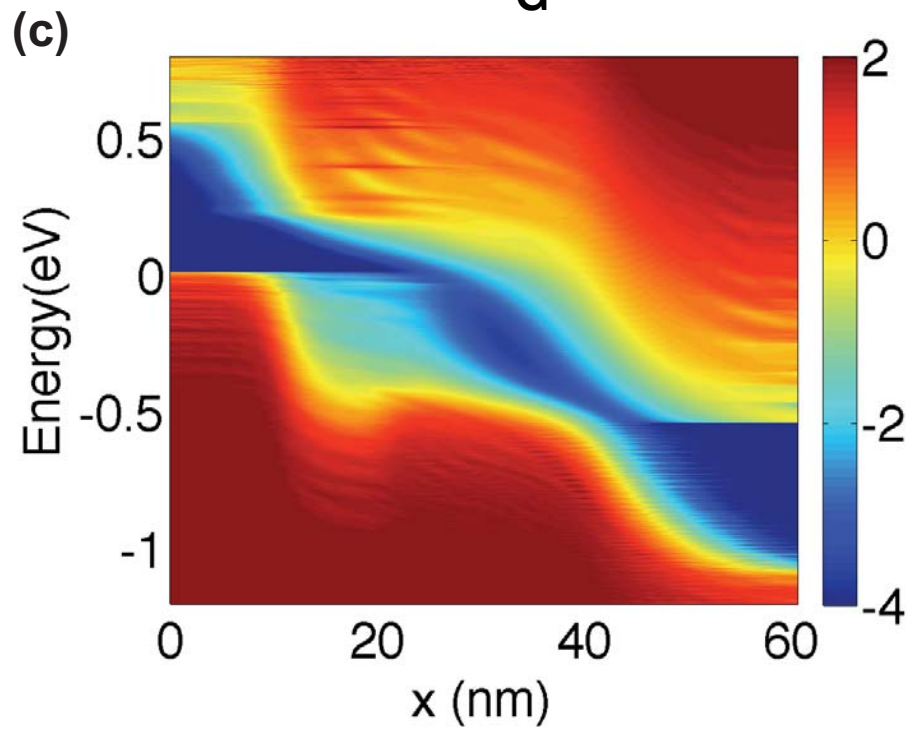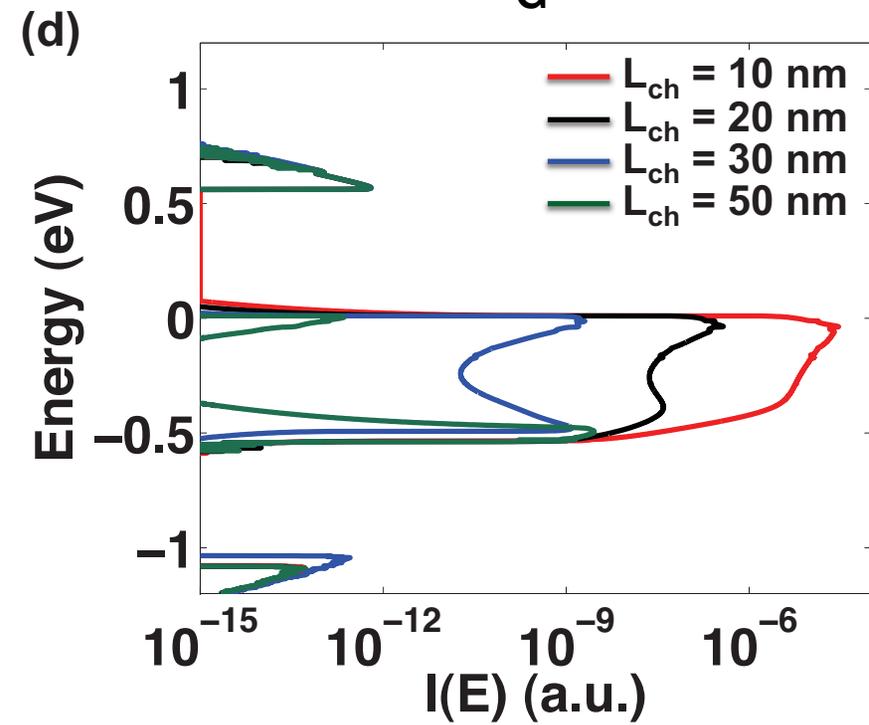